\documentclass[sigconf]{acmart}
\AtBeginDocument{%
  }

\copyrightyear{2026}
\acmYear{2026}
\setcopyright{cc}
\setcctype{by}
\acmConference[DAC '26]{63rd ACM/IEEE Design Automation Conference}{July 26--29, 2026}{Long Beach, CA, USA}
\acmBooktitle{63rd ACM/IEEE Design Automation Conference (DAC '26), July 26--29, 2026, Long Beach, CA, USA}
\acmDOI{10.1145/3770743.3803989}
\acmISBN{979-8-4007-2254-7/2026/07}

\usepackage[english]{babel}
\usepackage{blindtext}
\usepackage{xspace}
\usepackage{fancyhdr}
\usepackage{xcolor}
\usepackage{pifont}
\usepackage{xurl}
\usepackage{algorithm}
\usepackage{algpseudocode}
\usepackage{setspace}
\usepackage{amsmath}
\usepackage{multirow}
\usepackage{enumitem}
\usepackage{tikz}
\usepackage{multirow}
\usepackage{CJKutf8}
\usepackage{CJKfntef}
\usepackage{tabularx}
\usepackage{graphicx}
\usepackage{caption}
\usepackage{subcaption}
\usepackage{makecell} 
\newcommand{\name}{\textsc{Amoeba}\xspace}
\newcommand{\parabf}[1]{\noindent\textbf{#1}}

\newcommand{\eg}{e.g.,\xspace}
\newcommand{\ie}{i.e.,\xspace}
\long\def\com#1{}
\usepackage{xcolor}
\definecolor{darkgreen}{RGB}{0, 151, 0}
\definecolor{wine}{RGB}{128,0,0}

\algtext*{EndIf}
\algtext*{EndFor}
\algtext*{EndWhile}
\algtext*{EndFunction}
\algtext*{EndLoop}

\begin{document}

\title[\name]{\huge \name: Runtime Tensor Parallel Transformation for LLM Inference Services}

\author{
Haoyu Chen\textsuperscript{1},
Xue Li\textsuperscript{2},
Kun Qian\textsuperscript{2,*},
Yu Guan\textsuperscript{2},
Jin Zhao\textsuperscript{1,3,*},
Xin Wang\textsuperscript{1}
\and
\textsuperscript{1}Fudan University China,
\textsuperscript{2}Alibaba Group China,
\textsuperscript{3} Songshan Lab China,\\
\textsuperscript{*} Corresponding Authors: kunqian.qk@alibaba-inc.com, jzhao@fudan.edu.cn
}

% \author{Haoyu Chen}

% \affiliation{%
%   \institution{Fudan University}
%   % \city{Shanghai}
%   % \country{China}
% }

% \author{Xue Li}
% % \email{youli.lx@alibaba-inc.com}
% \affiliation{%
%   \institution{Alibaba Group}
%   % \city{Hangzhou}
%   % \state{Zhejiang}
%   % \country{China}
%   }

% \author{Kun Qian}
% % \email{kunqian.qk@alibaba-inc.com}
% \authornote{Corresponding Authors}
% \affiliation{%
%   \institution{Alibaba Group}
%   % \city{Hangzhou}
%   % \state{Zhejiang}
%   % \country{China}
% }

% \author{Yu Guan}
% % \email{guanyu.gy@alibaba-inc.com}
% \affiliation{%
%  \institution{Alibaba Group}
%  % \city{Beijing}
%  % \country{China}
%  }

% \author{Jin Zhao}
% % \email{jzhao@fudan.edu.cn}
% \affiliation{%
%   \institution{Fudan University and Songshan Lab}
%   % \city{Shanghai}
%   % \country{China}
%   }

% \author{Xin Wang}
% % \email{xinw@fudan.edu.cn}
% \affiliation{%
%   \institution{Fudan University}
%   % \city{Shanghai}
%   % \country{China}
%   }

\renewcommand{\shortauthors}{Haoyu et al.}

\maketitle

\section*{Abstract}

In Large Language Model (LLM) inference services, it is challenging to make a parallelism strategy configuration, to efficiently process the requests of variance context lengths. Requests of long context require high degree of parallelism to provide more memory for Key-Value (KV) Cache, while requests of short context prefer low degree of parallelism to increase concurrency, thus improving throughput.
%However, there is an intrinsic trade-off:
%while leveraging parallelism strategies, such as Tensor Parallelism (TP), can coordinate multiple GPUs to accommodate larger context lengths, it inevitably results in degraded overall throughput.
%Deploying multiple instances with varying parallelism configurations fails to effectively address dynamic workloads, potentially leading to low GPU resource utilization.

To maintain high throughput while supporting large context lengths on demand, we propose \name, a runtime Tensor Parallel (TP) transformation for online LLM inference services, which adaptively adjusts the TP of running instances to align with the dynamics of incoming requests.
Evaluations using real-world traces show that \name improves throughput by 1.75$\times$-6.57$\times$ compared to state-of-the-art solutions.

\section{Introduction}\label{sec:intro}

As large language models (LLMs) play increasingly important roles, the demand for LLM inference services in cloud environments has surged dramatically. 
Delivering optimal performance for inference requests has become a paramount requirement.
Significant efforts have been devoted to optimizing performance through instance-level acceleration mechanisms~\cite{Alpa, FlashAttention, FlashDecode, PageAttention,FlexGen,LightSeq,FastServe,E3} and global scheduling frameworks~\cite{Nexus, DVABatch, TritonServer, Shepherd, AlpaServe, Orca,Llumnix}.

Finding an optimal parallel strategy is one of the challenges in LLM inference services. In production workloads, requests are of variance context length. Requests with long context length require more GPU memory for KV Cache, so parallelized inference, like Tensor Parallel (TP), becomes mandatory to provide enough memory from multiple GPUs. However, parallelized inference increases the overhead from computation partitioning and inter-worker communication, leading to GPU utilization loss (\eg 57\% reduction from TP=1 to TP=4, based on our practical measurements). As a result, requests with short context length prefer no parallelism, which can serve more requests concurrently to improve the throughput.

A na\"ive solution is to provide multiple LLM serving instances with different parallelism strategy configurations (\eg 80\% instances with TP=1 and 20\% with TP=4). However, in production, the distribution of requests' context length is jittery and unpredictable (\S\ref{motiv:limitation}). Such a static TP setting compels
workers to operate in an inefficient way when long-context
requests are absent, or fails to serve some requests in the burst of long context requests.

Under dynamic request workloads, runtime parallel transformation is a promising technique to provide the ability to serve long-context requests while maintaining high throughput for short-context requests. KunServe extends Pipeline Parallel (PP) on demand without recomputing existing requests, and LoongServer extends Sequence Parallel (SP). Unfortunately, in productive online LLM inference services, due to serious performance degradation, only 3.2\% of instances apply PP or SP. Most instances (91.7\%) only apply TP, so they can not benefit from above solutions.

In this paper, we propose \name, the first runtime tensor parallel transformation for online LLM inference services. \name adapts the degree of TP on demand. For example, when long-context requests arrive, it merges four $TP1$ instances into a single $TP4$ instance, without recomputation of running requests. After long-context processing completes, the $TP4$ instance can be decomposed into four $TP1$ instances to maximize
throughput.

Different from other parallel strategies (\eg PP and SP), enabling runtime TP transformation is more challenging:
\begin{itemize}[leftmargin=*,nolistsep,topsep=0pt]
\item \textit{Challenge~1}: In TP transformation, KV Cache requires redistributing the KV Cache across GPUs in kv-head dimension, causing severe memory fragmentation which undermines the memory-saving benefits of the transformation. De-fragmentation introduces prohibitive overhead. 

\item \textit{Challenge~2}: In TP transformation, the model weight requires the division of the weights among GPUs. 
Page-wise memory management uses a minimum unit of 2MB~\cite{2MB_granularity}. 
However, the weight partitions of most models do not align perfectly with this 2MB granularity. 
This misalignment results in additional memory allocation and data movement.

\item \textit{Challenge~3}: How to schedule the TP transformation in dynamic workload?

\end{itemize}

To solve the above challenges, we (1) design a page-friendly header-centric KV Cache layout~(\S\ref{sec:subsec:KV_transformation}) that reduces memory overhead by 91.6\% and time cost by 86\% during KV Cache transformation, (2) by closely examining the partition and calculation patterns, we propose a parallelism-aware padding technique that allows in-place weight transformation (\S\ref{sec:subsec:weight_transformation}),
%We further conduct a series of optimizations and generalizations for our transformation solution.
and (3) we develop a TP transformation-aware scheduler that tightly integrates parallelism adaptation with request dispatching (\S\ref{sec:scheduler}). 

Our key contributions include:
\begin{itemize}[leftmargin=*,nolistsep,topsep=0pt]
\item A thorough analysis of the trade-off between performance and long-context support in LLM inference, highlighting the requirement and challenges of TP parallelism transformation (\S\ref{sec:motivation}).

\item The detailed design for the KV Cache transformation, the model weight transformation, layer-by-layer optimizations and the transformation aware scheduler (\S\ref{sec:design}).

%$\bullet$ The transformation-aware scheduler that dynamically adapts to workloads (\S\ref{sec:scheduler}).

\item Experiments demonstrate \name improving up to 6.57$\times$ throughput and decreasing 97\% transformation cost compared with the state-of-the-art alternatives
(\S\ref{sec:evaluation}).
\end{itemize}

\section{Motivation}\label{sec:motivation}
\vspace{-0.1in}
\begin{figure}[htp]
    \centering
    \includegraphics[width=\linewidth]{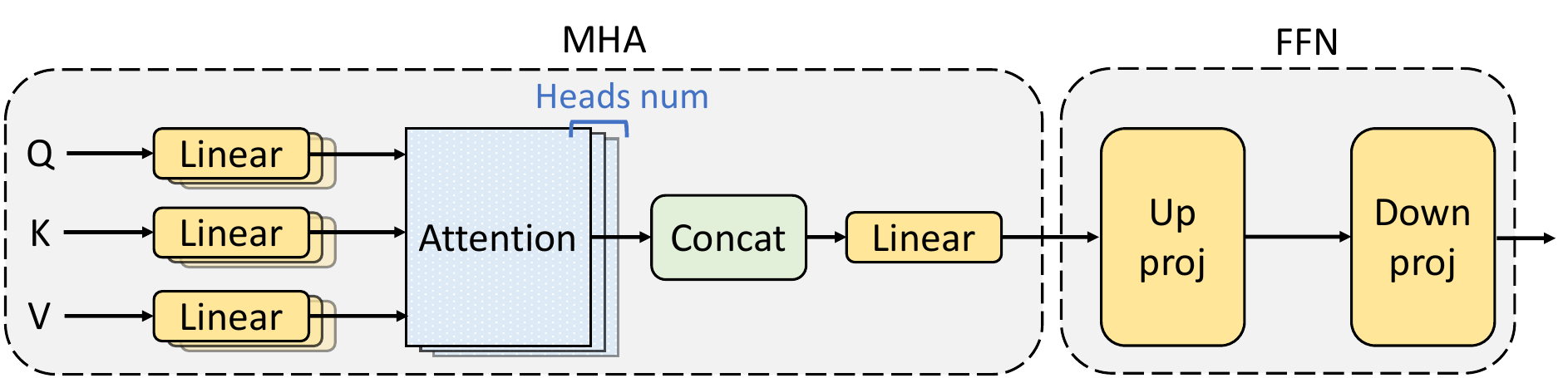}
    \vspace{-0.3in}
    \caption{Transformer layer overview.}
    \label{fig:background}
    \vspace{-0.3in}
\end{figure}

\subsection{LLM Serving}
\vspace{-0.05in}

An LLM inference service receives requests from remote clients and generates responses in a token-by-token way.
%The entire serving process for each request consists of two main phases: prefill and decode.
%The prefill phase processes the entire input request to construct the KV cache and generate the first token. 
%The decode phase recursively generates new tokens based on the KV cache of all previous tokens until it reaches the end-of-sequence (EOS) token or the maximum token length.
An large language model is constructed by a series of transformer layers, and the generation of each token requires passing through all layers.
As shown in Figure~\ref{fig:background}, a typical transformer layer consists of two main parts: Multi-Head Attention (MHA)\footnote{Group-Query Attention (GQA) is a variance of MHA, and we use MHA to refer to this family of attention mechanisms.} and Feed-Forward Network (FFN). 
The MHA computation requires the context of all previous tokens. To eliminate redundant calculations, KV Cache is used to store the internal results of these prior tokens. In contrast, the FFN is primarily constructed from two General Matrix Multiplications (GEMM), which necessitate fixed-size model weights.

\subsection{Supporting Long Context by Parallelism}
In LLM inference services, enlarging parallelism is the dominant technique to support long context without accuracy loss, because it utilizes multiple GPUs for a single service instance (\eg to accommodate larger KV Cache by increasing GPU memory). A series of parallelism strategies has been proposed.

The most common parallelism strategy is Tensor Parallel (TP).
In each transformer layer, different headers in the MHA are distributed across various workers  (each worker possesses a GPU), and the weights in the FFN are divided into segments, each managed by a different worker. 
During the computation of each layer, each worker computes partial results and uses AllReduce among workers to generate the final results. 
With TP, the weights of MHA, FFN, and KV Cache are evenly distributed across all workers.
%Different headers in MHA are evenly distributed to different workers, while the tensors in the MLP are evenly split into parts and maintained by different workers.
%During the calculation of each layer, each worker calculates partial results and leverage AllReduce among different GPUs to generate the entire results.
%With TP, weights of MHA, KV cache and weights of MLP can all be evenly distributed among all GPUs.
%Thus, all GPUs can be used concurrently to calculate each transformer layer.

Besides TP, other parallelism strategies are also introduced in the inference scenario.
Pipeline Parallel (PP) splits different transformer layers among different workers to divide memory pressure across different GPUs.
Sequence Parallel (SP) splits the input sequence into multiple parts and serves them with different GPUs.
For serving Mixtures-of-Experts (MoE) models, Expert Parallel (EP) is designed to optimize the use of more GPUs.
\vspace{-0.1in}
\subsection{Dilemma between Peak Throughput and Large Context}

While parallelism offers increased GPU memory for supporting long context, it is not a free lunch: the overall inference throughput decreases accordingly.

Taking TP for example, we conduct an experiment (serving Qwen2.5-32B with 4 H20 GPUs) to show the pros and cons introduced by TP.
The model size of Qwen2.5-32B (BF16 datatype) is 62.34 GB, runtime activations take 14.3 GB, and the GPU memory size of H20 is 96 GB.
Therefore, with 4 H20 GPUs, actually we have three typical deployment choices: $4\times (TP1)$, $2\times (TP2)$ and $TP4$.
%As shown in Figure~\ref{fig:4XTP1}, w
When deploying with $4\times (TP1)$, 64.9\% GPU memory is used to maintain model weights.
On the other hand, with $TP4$, 
%as shown in Figure~\ref{fig:1XTP4}, 
only 16.2\% is used to maintain model weights on each GPU.
As shown in Table~\ref{tab:TP_performance}, 
$TP1$ can only support a maximum context length of 3.75K, while $TP4$ can serve 32$\times$ larger.
On the other hand, serving the same workloads (context length of 1K tokens),
%with 1024 input sequence and 8 batch size, 
the performance varies according to different parallelism configurations.
$4\times (TP1)$ can deliver 233\% throughput compared to $TP4$, while keeping the target Service Level Objective (SLO), \eg Time to first token (TTFT) < 10s and Time per output token (TPOT) < 100ms. The reason is that high degree of TP requires extra TP communication.

\begin{table}[!t]
\centering
\small
\caption{Performance of different TP strategies.}
\vspace{-0.1in}
\label{tab:TP_performance}
\begin{tabular}{|c|c|c|c|}
\hline
\textbf{} & \textbf{TP1} & \textbf{TP2} & \textbf{TP4} \\ \hline \hline
% Maximal supported sequence & 30k & 330k & 964k \\ \hline
Maximal supported context length & 3.75K & 41.25K & 120.5K \\ \hline
%TTFT & 577 ms & 338 ms & 220 ms \\ \hline
%TPOT & 88 ms & 81 ms & 72 ms \\ \hline
Single instance throughput & 448 tps & 670 tps & 767 tps \\ \hline
Total throughput & 1792 tps & 1340 tps & 767 tps\\ \hline
\end{tabular}
\vspace{-0.1in}
\end{table}

\begin{figure}[!t]
    \centering
    \begin{subfigure}{0.42\linewidth}
	\centering
	\includegraphics[width=\linewidth]{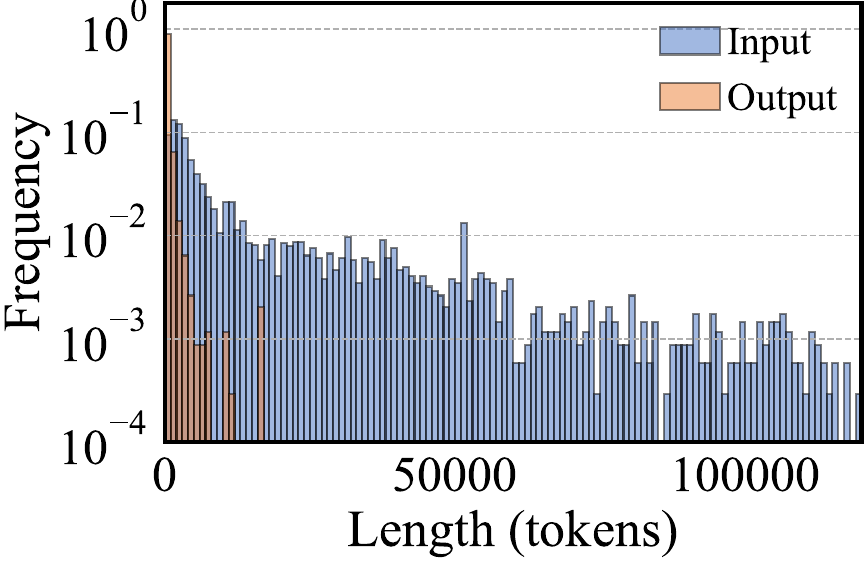}
	\vspace{-0.2in}
	\caption{Input/output length.}
    \vspace{-0.15in}
	\label{fig:length}
	\end{subfigure}
	\hfill
    \begin{subfigure}{0.47\linewidth}
	\centering
        \includegraphics[width=0.95\linewidth]{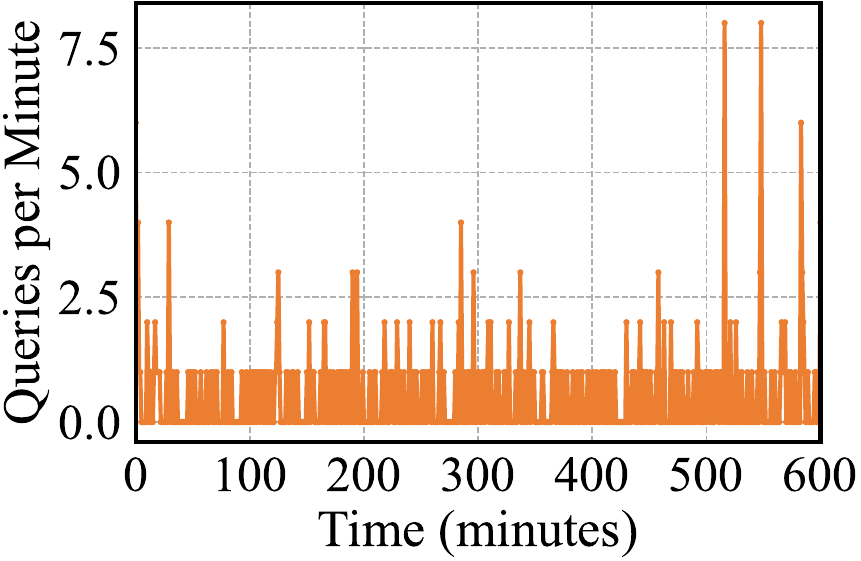}
        \vspace{-0.1in}
	\caption{Long requests' pattern.}
    \vspace{-0.15in}
	\label{fig:qps}
	\end{subfigure}
    \caption{Dynamic workload in LLM serving. }
    \vspace{-0.25in}
    \label{fig:workload}
\end{figure}

Similarly, the throughput degradation also happens in other parallelism implementations (\eg EP, SP, and PP).
This indicates that the optimal parallelism varies for different context lengths.
\vspace{-0.05in}
\subsection{Existing Solutions and Limitations}\label{motiv:limitation}
\vspace{-0.05in}
\parabf{Deploying instances of different parallelisms.}
The straight-forward solution is to deploy multiple instances of different parallelisms (\eg one $TP4$ and four $TP1$ instances on an 8-GPU host). While this solution is used in our production, it leads to inefficient utilization of GPUs, reducing the overall throughput.

We investigate statistics related to user requests in our production deployment. 
Figure~\ref{fig:length} shows the distribution of request input/output lengths for Qwen2.5-32B.
While requests with short context lengths constitute the majority of the workload, the length distribution exhibits an extremely long-tail property, indicating that 
$TP4$ deployment is essential in production. 
Additionally, we analyze the arrival pattern of long requests (which exceed the maximum supported sequence length of $TP2$ deployment) in Figure~\ref{fig:qps}, demonstrating that the traffic exhibits significant dynamics.
The above results reveal that long requests occur sporadically.
Therefore, reserving dedicated $TP4$ instances to accommodate these long requests is highly inefficient.

\parabf{Runtime parallel transformation}.
KunServe~\cite{KunServe} and LoongServe~\cite{LoongServe} provide dynamic PP and SP transformation for dynamic workloads.
However, both PP and SP severely reduce GPU utilization in online LLM inference. When $PP=N$, during the serving of each request, only $1/N$ GPUs are activated in any time slot, leading to low GPU utilization~\cite{TP_is_good1}.
As for $SP=N$, each worker hosts the entire model and more communication is required during MHA~\cite{LoongServe}.
Due to this drawback, TP is the dominant parallelism strategy in production LLM inference. According to our statistics (containing thousands of instances), 91.7\% of multi-GPU LLM serving instances employ TP.
3.2\% of instances employ TP+EP, while none employ PP/SP. As a result, most LLM services can not benefit from these solutions. 

Seesaw~\cite{Seesaw} try to migrate running requests to CPU shared memory, but memory offloading and loading cause up to 41$\times$ time cost according to our evaluations~(\S\ref{subsubsec:hybrid}).

\begin{figure}[!ht]
  \centering
  \includegraphics[width=\linewidth]{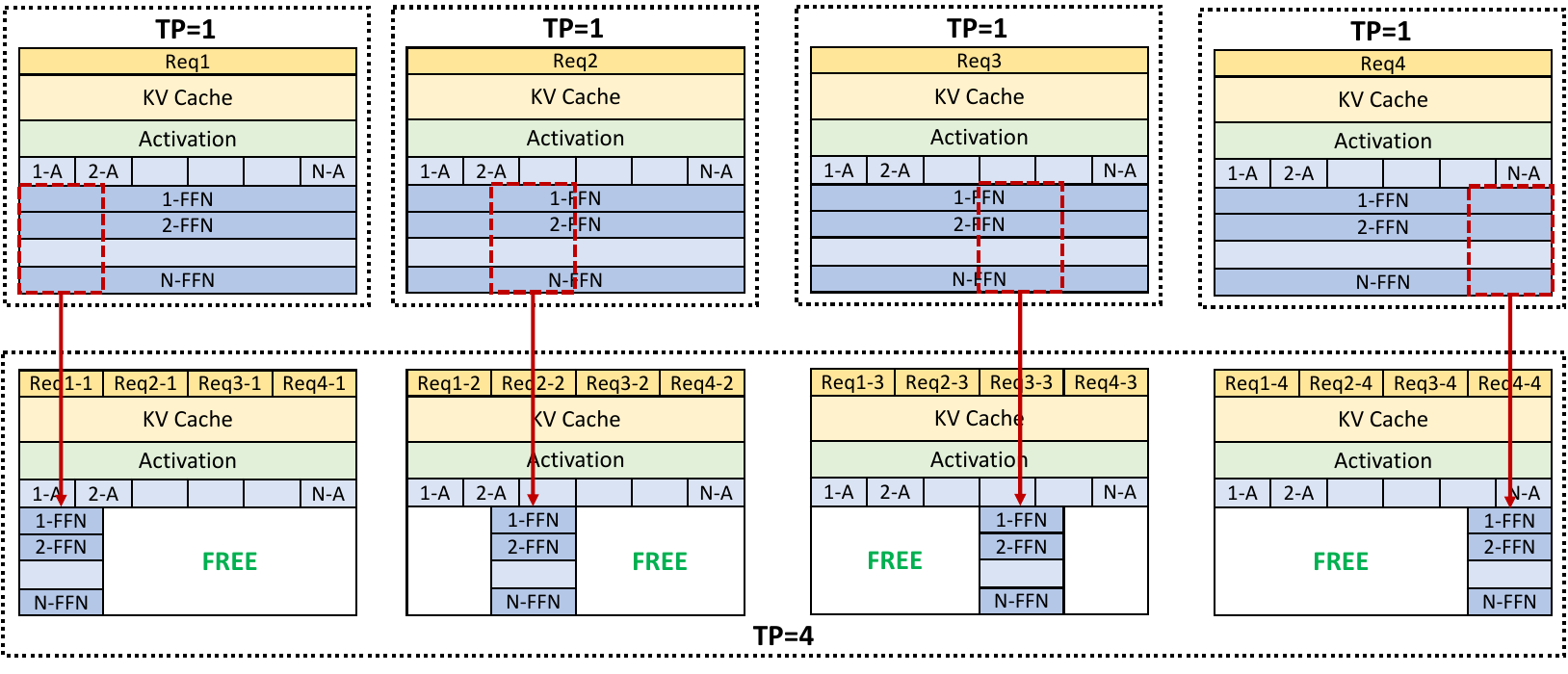}
  \vspace{-0.2in}
  \caption{Parallelism transformation overview from $TP1$ to $TP4$.}
  \vspace{-0.2in}
    \label{fig:basic_idea}
\end{figure}

\vspace{-0.05in}
\section{\name Design}\label{sec:design}

We present \name, the first runtime tensor parallel transformation for online LLM inference. The core idea is as follows: 
By default, all instances operate with $TP1$ for optimal throughput.
When a request exceeds the context length limit of $TP1$, multiple GPUs on the same host are dynamically aggregated into a $TP2$ or $TP4$ instance (Figure~\ref{fig:basic_idea}) to transfer memory space from FFN to KV Cache.
After serving these long requests, the system can revert to $TP1$ to maximize overall throughput. Since TP is dominant in parallel strategies of LLM serving, \name can be easily deployed in production.

\name includes three design choices: \S\ref{sec:subsec:KV_transformation} proposes a page-friendly header-centric KV Cache layout to reduce memory overhead and cudaMemcopy operations in KV Cache transformation (\textit{Challenge~1} in \S\ref{sec:intro}). \S\ref{sec:subsec:weight_transformation} proposes a parallelism-aware padding technique that allows
in-place weight transformation to avoid weight migration (\textit{Challenge~2}). \S\ref{sec:scheduler} proposes a scheduler to control aggregation and split of TP (\textit{Challenge~3}).

\vspace{-0.05in}
\subsection{Transformation of KV Cache}
\label{sec:subsec:KV_transformation}
\vspace{-0.05in}

\begin{figure}[t]
  \centering
  \includegraphics[width=0.9\linewidth]{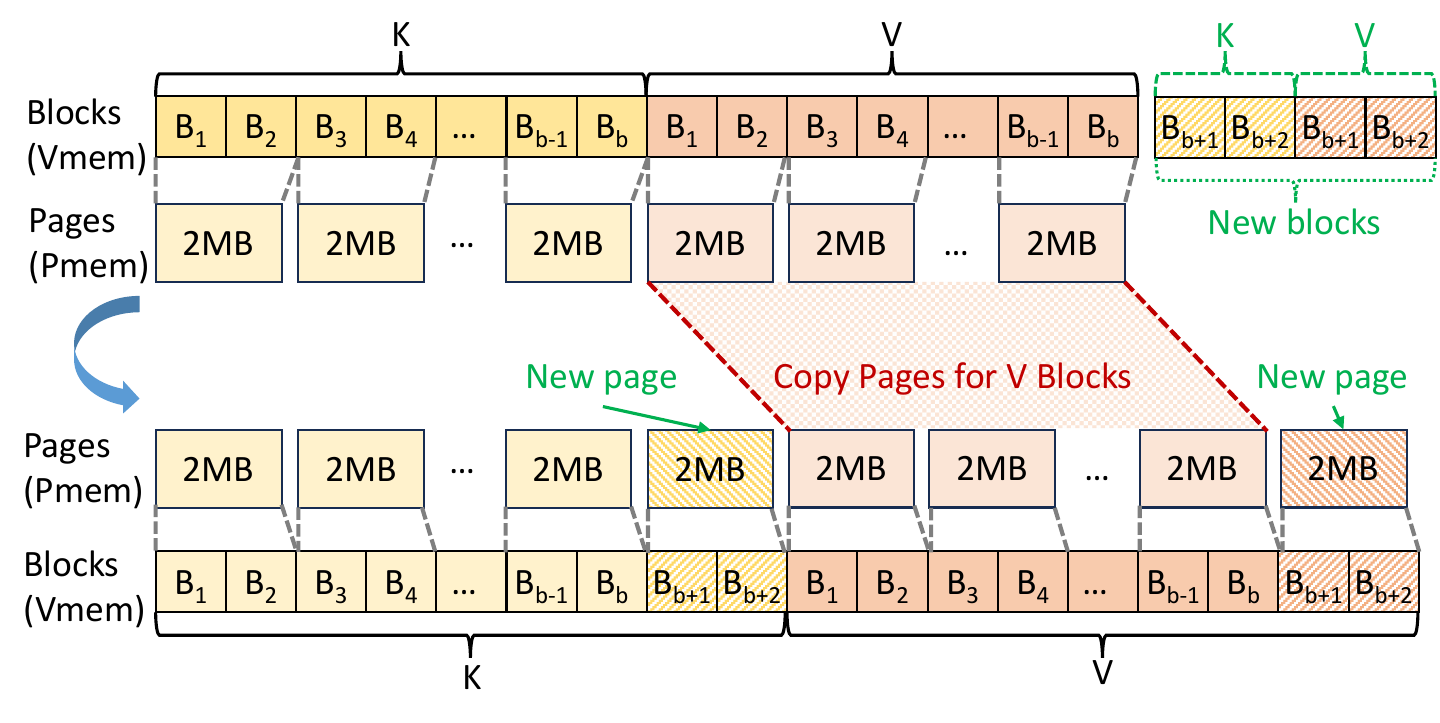}
  \vspace{-0.2in}
  \caption{Expand KV Cache with page-wise memory management.}
  \vspace{-0.2in}
    \label{fig:kv_cache_0}
\end{figure}

Transformation of KV Cache introduces huge overload of memory management. The overload primarily consists of two parts.

The first overload lies in cudaMemcopy operations introduced by page-wise memory management.
To enable flexible KV Cache expansion during transformation, we utilize page-wise memory management~\cite{2MB_granularity} to store the KV Cache. 
However, the standard KV Cache layout in mainstream inference engines~\cite{vLLM, SGLang} is incompatible with flexible expansion, because it requires K and V values to reside in two contiguous blocks. 
As illustrated in Figure~\ref{fig:kv_cache_0}, adding new pages necessitates additional shifting operations to V values (via copy or memory unmap+remap operations) to maintain contiguous allocation of the entire K and V.

\begin{figure*}[htbp]
    \centering
    \begin{subfigure}{0.24\linewidth}
	\includegraphics[width=\linewidth]{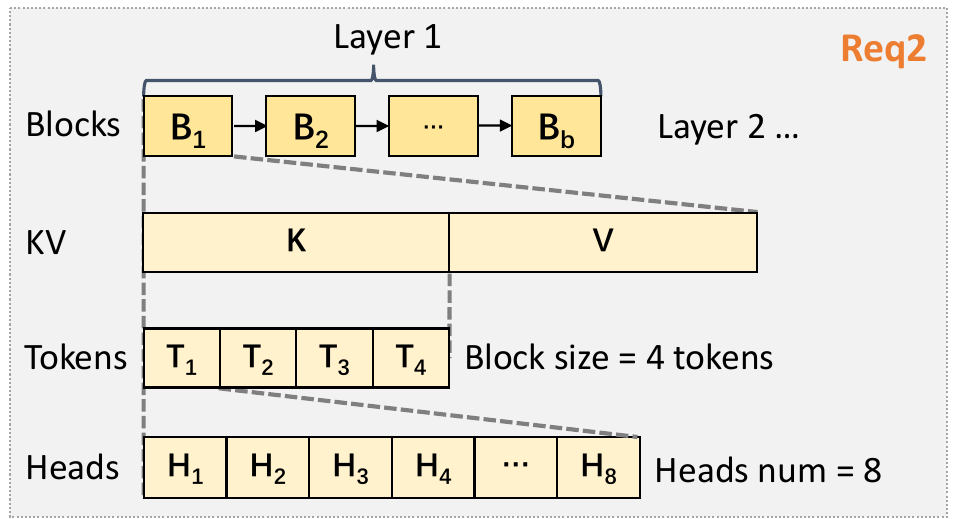}
	\caption{Init status.}
\vspace{-0.1in}
	\label{fig:KV_cache_1}
	\end{subfigure}
    \hfill
    \begin{subfigure}{0.49\linewidth}
	\includegraphics[width=\linewidth]{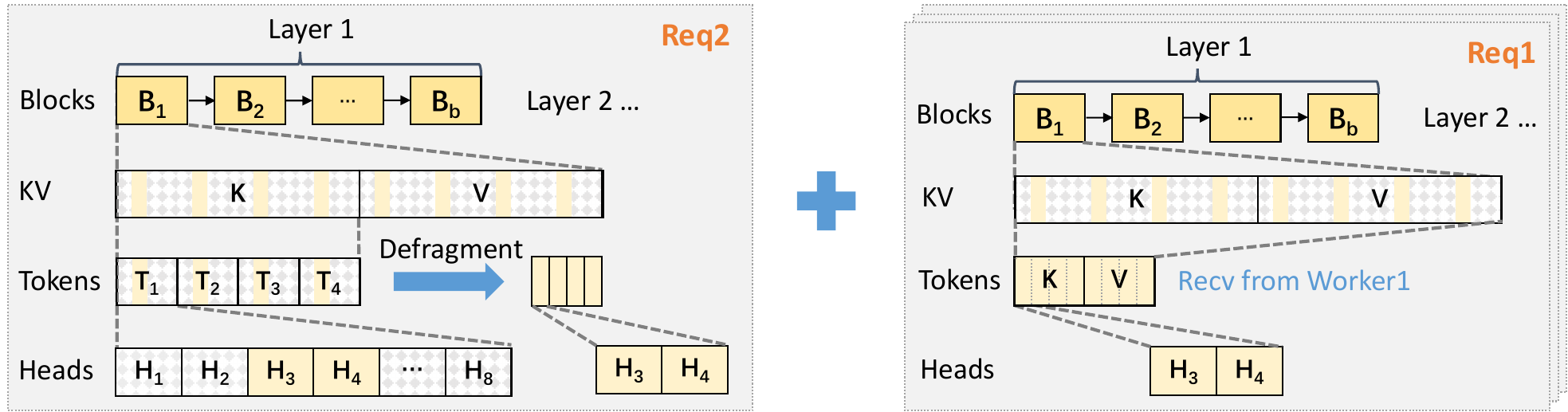}
	\caption{KV Cache migration and de-fragmentation.}
	\vspace{-0.1in}
	\label{fig:KV_cache_2}
    	\end{subfigure}
    \hfill
    \begin{subfigure}{0.24\linewidth}
	\includegraphics[width=\linewidth]{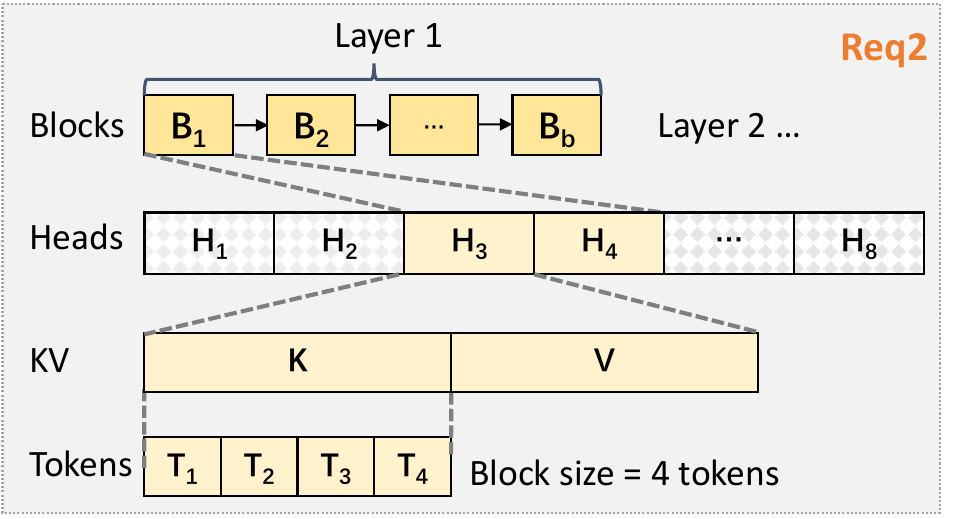}
	\caption{Optimized KV Cache layout.}
	\vspace{-0.1in}
	\label{fig:KV_cache_3}
	\end{subfigure}
	\hfill
    \caption{KV Cache migration solutions.}
    \vspace{-0.2in}
    \label{fig:KV_cache_fig}
\end{figure*}

The second overload lies in utilizing the memory with bubbles after transformation.
Figure~\ref{fig:KV_cache_fig} shows an example of KV Cache migration.
Each worker starts at $TP1$, and maintains the full KV Cache for requests it serves, we call them \textit{local requests} (Figure~\ref{fig:KV_cache_1}).
When transforming from $4\times (TP1)$ to $TP4$, for each layer, the KV Cache for each token must be divided and distributed among workers according to the number of headers in MHA.
As illustrated in Figure~\ref{fig:KV_cache_2}, this makes a KV Cache ``full of bubbles'', creating significant challenges for reusing the released memory space.

\begin{figure}[htbp]
  \centering
\includegraphics[width=\linewidth]{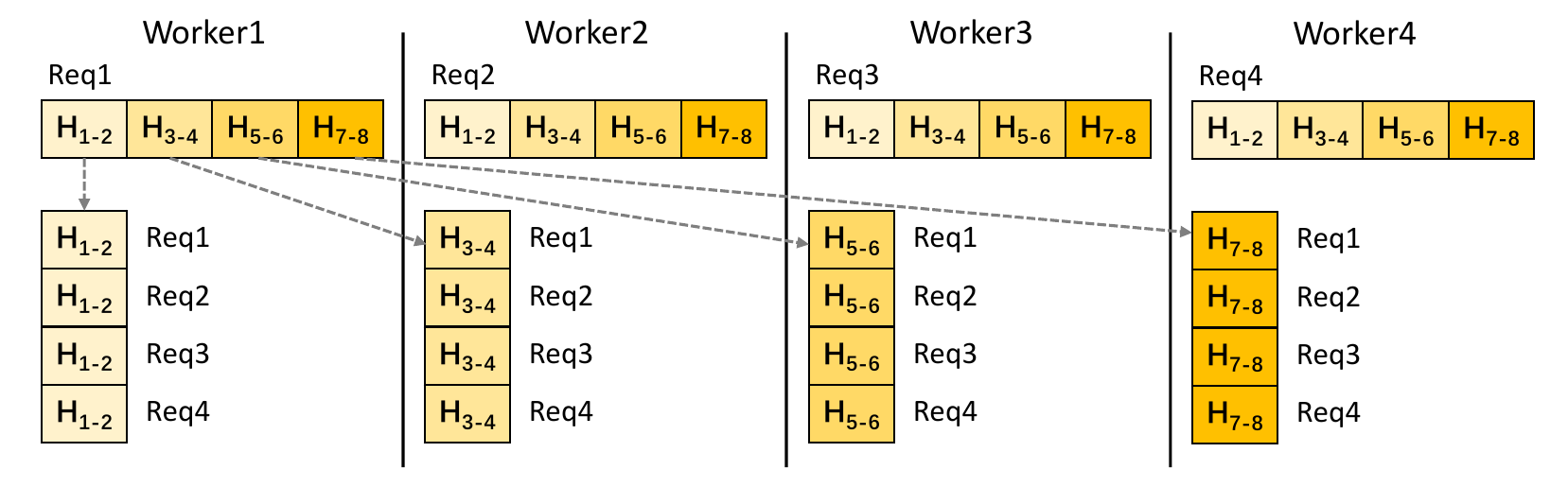}
\vspace{-0.2in}
	\caption{All-to-all KV Cache migration in head dimension.}
	\vspace{-0.15in}
	\label{fig:KV_cache_4}
\end{figure}

\parabf{Base solution: Migration and defragmentation based on page-friendly KV Cache layout.}
We redisign the KV Cache architecture to a page-friendly one by reorganizing the hierarchical relationship between K/V tensors and memory blocks.
The key innovation is arranging KV Cache for consecutive tokens to occupy adjacent blocks, aligning with the page allocation patterns. 
Thus, each new page can be directly attached to the end of the existing KV Cache with no shifting overhead.
% Thus, this design could eliminate the shifting overhead introduced by page-wise management.

To fully utilize memory bubbles after transformation, we utilize the released KV Cache space by de-fragmenting the local KV Cache to a more compact style after migration.
% The straightforward solution involves executing KV cache migration followed by  trimming local KV cache to a more compact style.
During migration, worker $W_i$ selectively remains headers in the range $(H/TP)*(i-1) + 1 \sim (H/TP)*(i-1) + H/TP$ for each locally stored token, where $H$ is the number of headers in MHA and $TP$ is the target TP configuration size.
Each worker sends the remaining headers to the corresponding workers.
In Figure~\ref{fig:KV_cache_2}, assuming $H=8$ and a $TP4$ configuration, 
$W_2$ retains $H3$ and $H4$, and sends $(H1+H2)/(H5+H6)/(H7+H8)$ to $W_1/W_3/W_4$, respectively.
Simultaneously, each worker prepares reserved pages to store the KV Cache received from remote requests.

\parabf{Improvement~1: In-place migration with header-centric KV Cache layout.}
Since the KV Cache for \textit{local requests} (\eg $Req2$ in Figure~\ref{fig:KV_cache_2}) is full of bubbles, we need to de-fragment the KV Cache to utilize the memory.
This de-fragmentation operation requires extensive copying, which significantly affects performance.
Our evaluations show that this leads to 12$\times$ memory usage and 2.6$\times$ processing time.
To avoid the KV Cache de-fragmentation, we find that non-contiguous memory spaces are caused by the token-first KV Cache layout (\ie token-level contiguous storage of different K/V pairs).
Since increasing TP configuration splits attention heads across workers (\eg 8 heads evenly distributed across 4 workers in $TP4$), we propose a {\it header-centric layout} that organizes K/V storage in the order of [Block, Header, K/V, Token]. 
As illustrated in Figure~\ref{fig:KV_cache_3}, this reorganization allows each block to generate compact memory segments during migration (Figure~\ref{fig:KV_cache_4}), allowing memory bubbles to be directly reused through block reshaping.

To enable the existing libraries to adapt to these novel KV Cache layouts, we developed a \texttt{kv\_stride\_order()} function for the attention kernel,
which automatically maps different layouts to corresponding element offsets and strides for the kernel.
Thus, from the attention kernel's perspective, the input remains unchanged, eliminating the need for further modifications.

\parabf{Improvement~2: Pipelining KV Cache transformation with existing GPU kernel computations.}
In KV Cache transformation, main GPU driver calls are  \texttt{cuMemMap} and \texttt{cuMemSetAccess}.
All of them are GPU driver functions that can run in parallel with GPU kernels. 
KV Cache transformation also require All-to-all communication which needs GPU SMs to achieve good performance. It would conflict with normal kernel executions.
Our strategy is to launch All-to-all on an independent communication stream, which will actually be executed when enough free SMs are available.

%\subsubsection{Overlapping with computation.}

\begin{figure}[!t]
    \centering
    \begin{subfigure}{\linewidth}
	\centering
	\includegraphics[width=\linewidth]{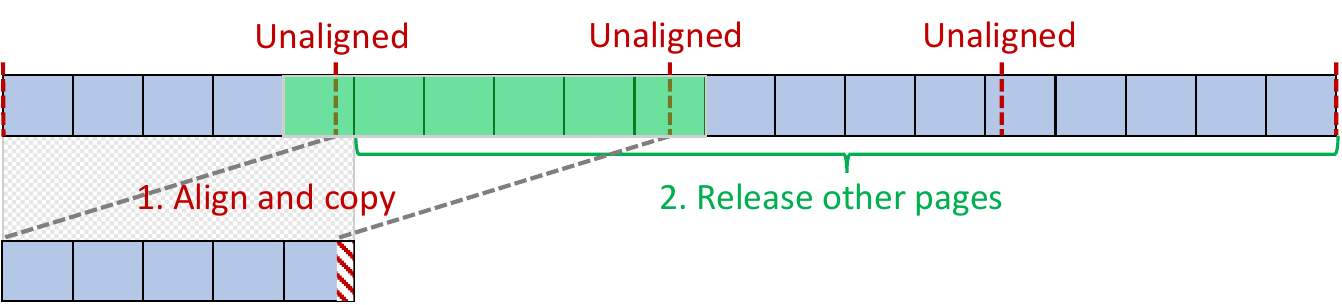}

	\caption{Base solution: partial swap weight migration.}
	\label{fig:block_fig_2}
	\end{subfigure}
    \begin{subfigure}{\linewidth}
	\centering
	\includegraphics[width=\linewidth]{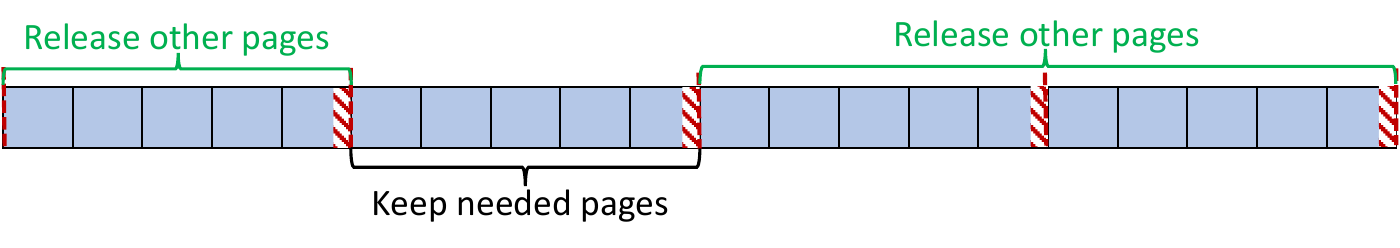}

	\caption{Improvement~1: in-place migration with padding.}
	\vspace{-0.15in}
	\label{fig:block_fig_3}
	\end{subfigure}

    \caption{One layer's model weight layout and migration solution.}
    \vspace{-0.3in}
    \label{fig:block_fig}

\end{figure}
\vspace{-0.1in}
\begin{table}[htbp]
\centering
\caption{FFN weight size in different models. Decimals mean unaligned placements of tensors.}
\vspace{-0.1in}
\footnotesize
\label{tab:mlp_weight}
\begin{tabular}{|c|c|c|c|}
\hline
\textbf{Model} & \textbf{Model Structure} & \textbf{\#Pages/Tensor} & \textbf{\#Pages/Tensor} \\ 
 & \textbf{[$H, I, \#Exp$]} & ($TP1$) & ($TP4$) \\ 
\hline \hline
GPT-OSS-120B & [2880, 2880, 128] & \textcolor{wine}{1012.5}/2025 & \textcolor{wine}{253.125}/\textcolor{wine}{506.25} \\
\hline
GPT-OSS-20B & [2880, 2880, 32] & \textcolor{wine}{253.125}/\textcolor{wine}{506.25} & \textcolor{wine}{63.281}/\textcolor{wine}{126.563} \\
\hline
Llama-3.1-70B & [8192, 28672, -] & 224 & 56 \\ \hline
Qwen2.5-32B & [5120, 27648, -] & 135 & \textcolor{wine}{33.75} \\ \hline
\end{tabular}
\vspace{-0.2in}
\end{table}

\vspace{-0.05in}
\subsection{Transformation of Model Weights}
\label{sec:subsec:weight_transformation}

In TP transformation of model weights, we focus on FFN weights, which constitute 88\% of the overall model weight, while keeping other weights duplicated for implementation simplicity. 

Transforming FFN weights from $TP1$ to $TP4$ may generate fragmented memory pages which cannot be efficiently utilized. As demonstrated in Figure~\ref{fig:block_fig_2}, a blue block refers to the minimum memory allocation page of 2MB enforced by CUDA's native memory management. While the red lines indicates the vertical partition boundaries of the FFN weight during transformation. 
In many cases, the weight partition boundary is not in align with page boundary.
If we directly remain the pages containing the weight partition (green part in Figure~\ref{fig:block_fig_2}) and free others.
The FFN cannot even been executed as the calcultion of GEMM requires weights stored in an aligned way. 
Unfortunately, our analysis of several representative open-source models, as shown in Table~\ref{tab:mlp_weight}, reveals that more than half of the models encounter this fragmentation issue.

\parabf{Base solution: Partial swap for model weight.}
To handle this problem, %a basic solution is to 
we conduct a two phase transformation as illustrated in Figure~\ref{fig:block_fig_2}, 
(1) we align the needed weight partition by moving it to the beginning of current layer's weight space through GPU driver operation \texttt{cudaMemcopy}
%(1) we align the weight partition (memory space between two consecutive red lines) to the first memory page and utilize GPU driver operation \texttt{cudaMemcopy} to move the entire partition 
and (2) we release other unused pages through \texttt{cuMemUnmap} operation.
% we conduct GPU driver operation \texttt{cudaMemcopy} 
%or employ a customized GPU kernel 
% to swap unaligned model weights for aligned ones (Figure~\ref{fig:block_fig_2}). 
% The required weight split is moved and aligned to the head of the allocated memory pages and the redundant pages are then released through cuMemUnmap operation.
%After copying and shifting the necessary weights, the previously used memory can be released.
However, the first phase results in extra weight copying that will block the CUDA stream.
%(1) additional memory usage and (2) extra copying of weights. 
%According to our measurement, these shifting and copy operations contribute \todo{XX\%} of the entire transformation.

\parabf{Improvement~1: In-place migration with weight padding.}
In parallelism transformation, performing any additional memory copies or movements on model weights leads to substantial time overhead.
However, our analysis (Table~\ref{tab:mlp_weight}) reveals that significant copying and shifting cost is triggered by small alignment deviations (\eg less than 0.7\% of the total weight size).

This observation motivates our {\it pre-alignment optimization strategy}: by introducing small padding to the weights, we can eliminate expensive runtime redistribution during transformation. 
Specifically, we proactively add padding at potential partitioning boundaries to ensure that the subsequent weights align with CUDA allocation granularity. 
Since the set of possible TP configurations (\eg $TP1/2/4$) is fixed for a given model, these partitioning boundaries can be predetermined during model loading.
With this pre-set padding, parallelism transformation can directly release unused pages, completely eliminating the weight copying overhead.
%identified in {\bf Challenge-2}. 

\begin{figure}[t]
  \centering
  \includegraphics[width=\linewidth]{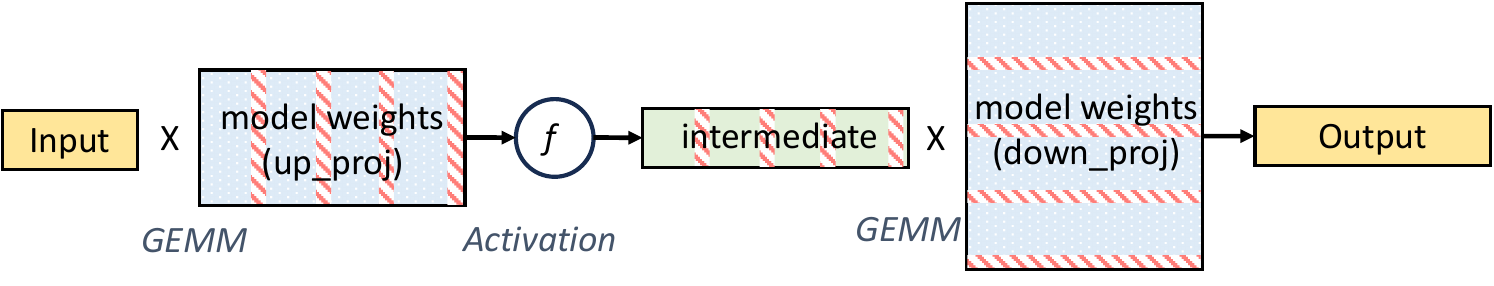}
  \vspace{-0.2in}
  \caption{FFN main workflow with padding.}
    \label{fig:ffn}
    \vspace{-0.28in}
\end{figure}

Now we prove that such a padding does not influence the computation result.
Figure~\ref{fig:ffn} illustrated the workflow of FFN implementation.
The raw FFN calculation can be presented as follows: 
\vspace{-0.15in}
\begin{equation}
    FFN = f(I\times U)\times D
\label{raw_MLP}
\end{equation}
where $I$ is the input tensor, $U$ is the \texttt{up\_proj}, $D$ is the \texttt{down\_proj}, and $f$ is the activation function.

We introduce even column-wise padding in $U$ and row-wise padding in $D$.
The padded $U'=[U_1,0,U_2,0,U_3,0,U_4,0]$, and $D'=[D_1^T, 0, D_2^T, 0, D_3^T,0, D_4^T,0]^T$.
After padding, the computation result $FFN'$ is:
\begin{equation}
\begin{split}
FFN' &=f(I\times U')\times D'\\
% &=f([I\times U_1,0,I\times U_2,0,I\times U_3,0,I\times U_4,0])\times D'\\
&= [f(I\times U_1),0,f(I\times U_2),0,f(I\times U_3),0,f(I\times U_4),0]\\
&\times [D_1^T, 0, D_2^T, 0, D_3^T,0, D_4^T,0]^T\\
&=f(I\times U)\times D = FFN
\end{split}
\label{eq:FFN'}
\end{equation}
Since $FFN=FFN'$, with our proposed even column-wise padding, we do not need further tensor movement or reshaping.

\parabf{Improvement~2: Pipelining CUDA Driver API Calls with computation kernels.}
During the parallelism scale-up, each worker only needs to free the memory pages through cuMemUnmap, which can be completely overlapped.
During the parallelism scale-down, All-to-algl communication is needed.
Being similar to \S\ref{sec:subsec:KV_transformation}, we utilize an independent communication stream to launch All-to-all kernels when enough SMs are available, thereby minimizing overhead.

%Through an analysis of TP execution patterns, we find that the entire model does not execute fully in parallel. 
%At the end of each computation module (MHA and MLP), a TP group must perform an all-reduce to aggregate computation results across workers before proceeding to the next execution stage. 
%This indicates that the finest granularity of TP execution and parallelism transformation operates at the computation module level.

\parabf{Discussion: Supporting EP transformation.}
The above design can also smoothly support EP transformation with minor change.
In MoE models, the FFN is stored in an expert-wise layout, allowing us to directly free unused expert weights.
We only need to determine how to release weights according to the LLM types.
In \S\ref{subsec:microbenchmark}, we involve Qwen3-30B-A3B to evaluate EP transformation.

\vspace{-0.1in}
\subsection{Transformation-Aware Scheduler}
\label{sec:scheduler}
\vspace{-0.05in}
We present a scheduler to control whether to execute TP transformation scale-up 
or scale-down  
based on dynamic requests. We also present the order of KV Cache and FFN to make transformation.

\parabf{Transformation scale-up.}
Once a new request arrival, the scheduler first identifies candidate instances capable of serving this request based on its input length and the current load (\eg KV Cache capacity) of each instance. 
If any available instance can serve this request, this request is scheduled to the available least-loaded instance.
If no available instance can support the input length, a parallelism scale-up is triggered on the least-loaded instances with same parallelism configuration.
The scale-up step size is set to $\times2$ (\eg from $TP1$ to $TP2$ or from $TP2$ to $TP4$), which delivers optimal performance in practice.
When consecutive long requests occur, the scheduler prioritizes instances already operating in higher parallelism degree to minimize the number of transformations.

During scale-up, \name first transforms FFN to free the memory occupation, and then for KV Cache.

\parabf{Transformation scale-down.}
Once long requests are completed and if any instances with $TP>1$ remain, the scheduler sets these instances as temporally unavailable for scheduling. 
Once sufficient free KV Cache capacity is available (50\% in practice), the scheduler will trigger a parallelism scale-down to maximize overall throughput.
Then these instances are set back to being available for scheduling.
We do not involve additional complex mechanisms for predicting the arrival of future long requests, and real-trace experiments (\S\ref{subsec:e2e}) validate the current scheduler is efficient in practice.

During scale-down, \name first makes parallelisam transformation for KV Cache to free the memory occupation, then FFN is able to gather the entire model weights for the next decoding step.
\begin{table}[!t]
\centering
\footnotesize
\caption{Details of selected models.}
\vspace{-0.1in}
\label{tab:models}
\begin{tabular}{|c|c|c|c|}
\hline
\textbf{Model} & \textbf{Parallelism} & \textbf{Weights Size} & \textbf{GPU Type} \\ \hline \hline
% Maximal supported sequence & 30k & 330k & 964k \\ \hline
Llama2-7B (BF16) & TP & 15.67 GB & A100 (40GB) \\ 
Llama3-8B (BF16) & TP & 16.66 GB & A100 (40GB) \\ 
Qwen2.5-32B (BF16) & TP & 62.34 GB & H20 (96GB) \\ 
Qwen3-32B (BF16) & TP & 62.34 GB & H20 (96GB) \\ 
Qwen3-30B-A3B (FP16) & TP+EP & 56.87 GB & H20 (96GB) \\
\hline
\end{tabular}

\end{table}

\vspace{-0.1in}
\section{Evaluation}\label{sec:evaluation}

In this section we show that:

$\bullet$ All techniques introduced by \name-i1 lead to less memory overhead and faster transformation speed (\S\ref{subsec:microbenchmark}).

$\bullet$ In real-world request workload, \name-i2 outperforms state-of-the-art solutions by 1.75x-6.57x in throughput (\S\ref{subsec:e2e}). 
% \end{itemize}

\begin{figure}[!t]
    \centering
    \begin{subfigure}{.49\linewidth}
	\centering
	\includegraphics[width=\linewidth]{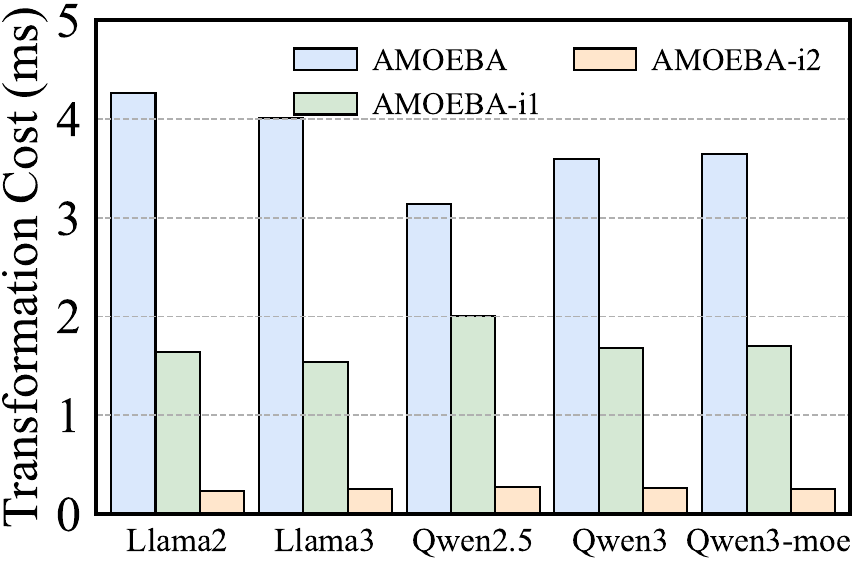}
        \caption{Transformation cost.}
        \label{fig:KV_cost_evaluation}
	\end{subfigure}
     \begin{subfigure}{.49\linewidth}
	\centering
	\includegraphics[width=\linewidth]{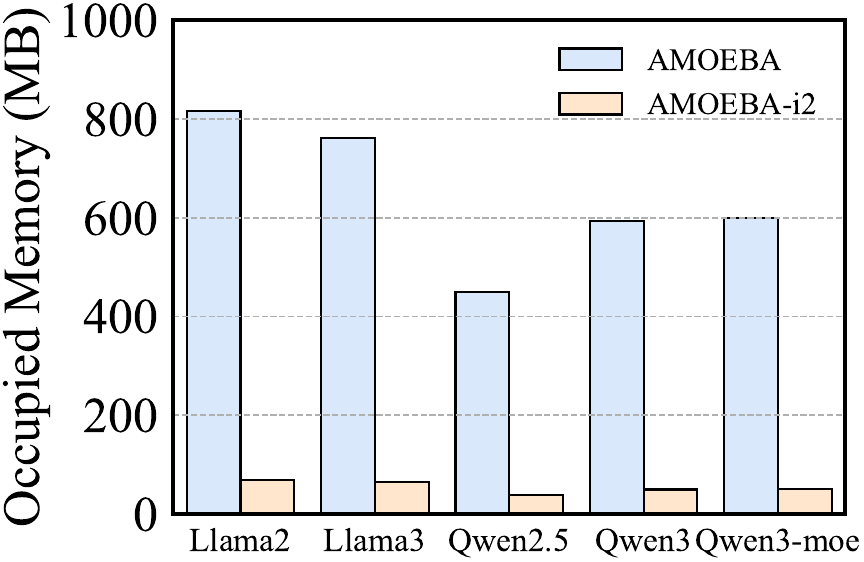}
	\caption{Occupied memory.}
        \label{fig:KV_memory_evaluation}
	\end{subfigure}
 \vspace{-0.15in}
    \caption{KV Cache transformation.}
    \vspace{-0.1in}
    \label{fig:KV_cache_evaluation}
\end{figure}
\begin{figure}[!t]
    \centering
    \begin{subfigure}{.49\linewidth}
	\centering
	\includegraphics[width=\linewidth]{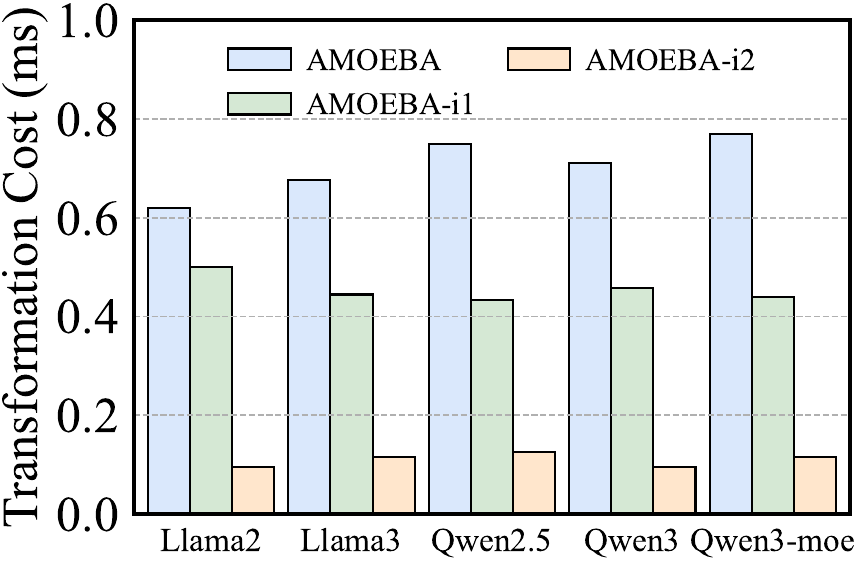}
        \caption{Transformation cost.}
        \label{fig:Model_cost_evaluation}
	\end{subfigure}
     \begin{subfigure}{.49\linewidth}
	\centering
	\includegraphics[width=\linewidth]{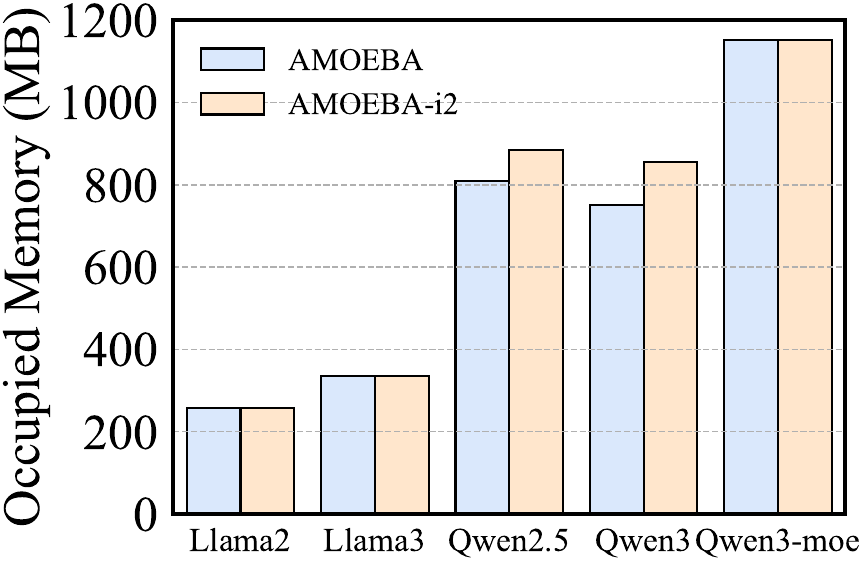}
	\caption{Occupied memory.}
        \label{fig:Model_memory_evaluation}
	\end{subfigure}
 \vspace{-0.1in}
    \caption{Model weights transformation.}
    \label{fig:model_weight_evaluation}
    \vspace{-0.2in}
\end{figure}

\begin{figure}[!t]
    \centering
     \begin{subfigure}{\linewidth}
	\centering
	\includegraphics[width=\linewidth]{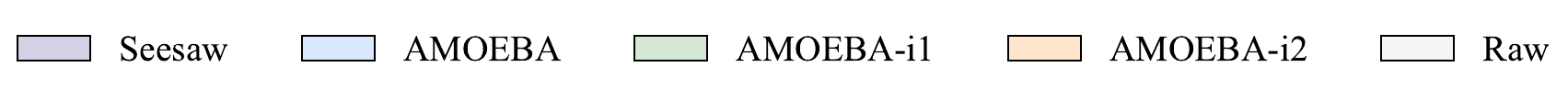}
	\end{subfigure}
    \begin{subfigure}{.49\linewidth}
	\centering
	\includegraphics[width=\linewidth]{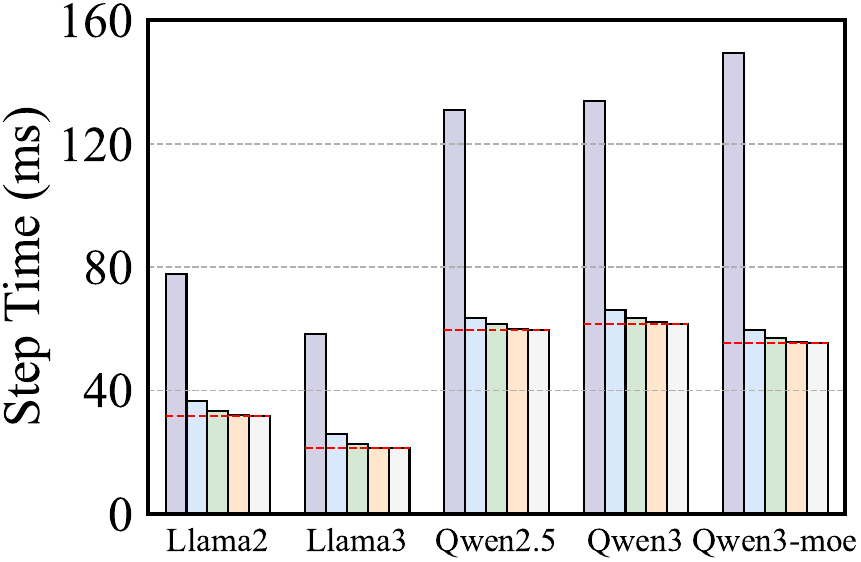}
        \caption{One layer.}
        \label{fig:overlap_benefit_1}
	\end{subfigure}
     \begin{subfigure}{.49\linewidth}
	\centering
	\includegraphics[width=\linewidth]{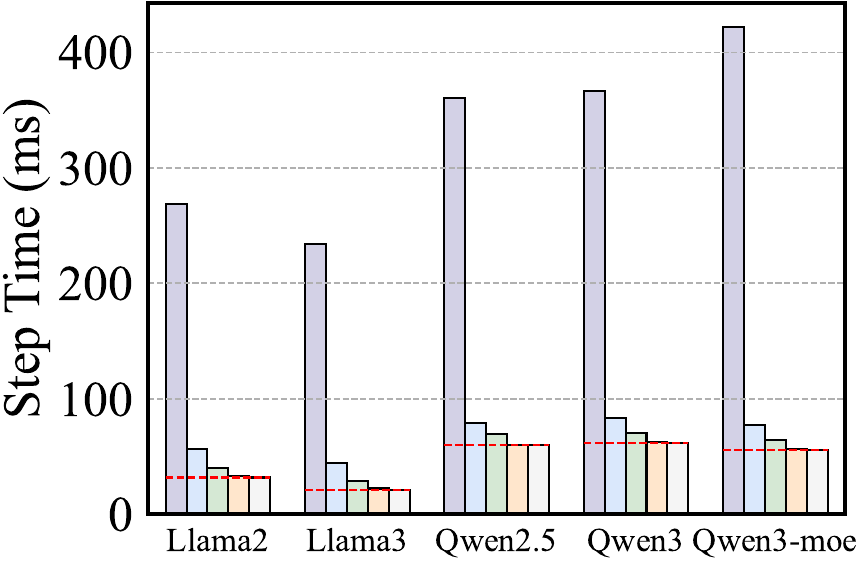}
	\caption{Five layers.}
        \label{fig:overlap_benefit_5}
	\end{subfigure}
    
    %\vspace{0.2in}
    
    \begin{subfigure}{.49\linewidth}
	\centering
	\includegraphics[width=\linewidth]{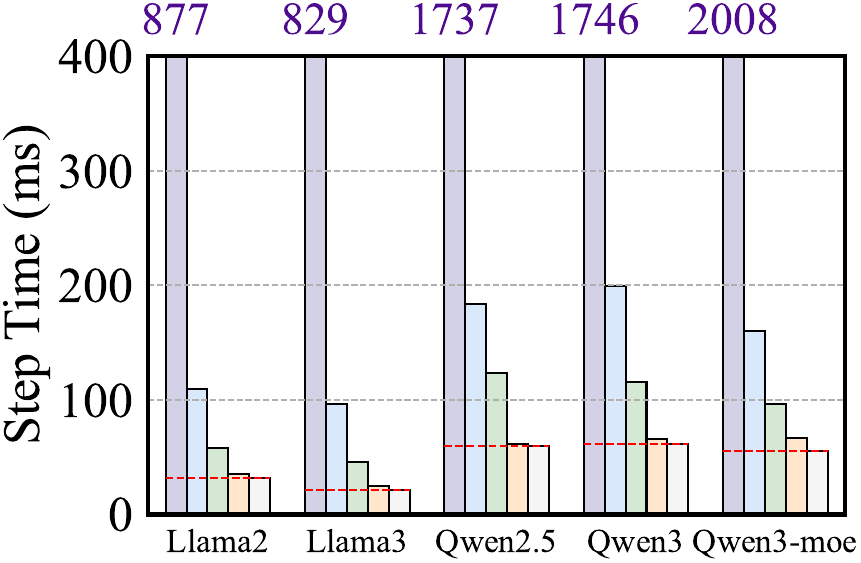}
        \caption{Half of all layers.}
        \label{fig:overlap_benefit_layer_2}
	\end{subfigure}
     \begin{subfigure}{.49\linewidth}
	\centering
	\includegraphics[width=\linewidth]{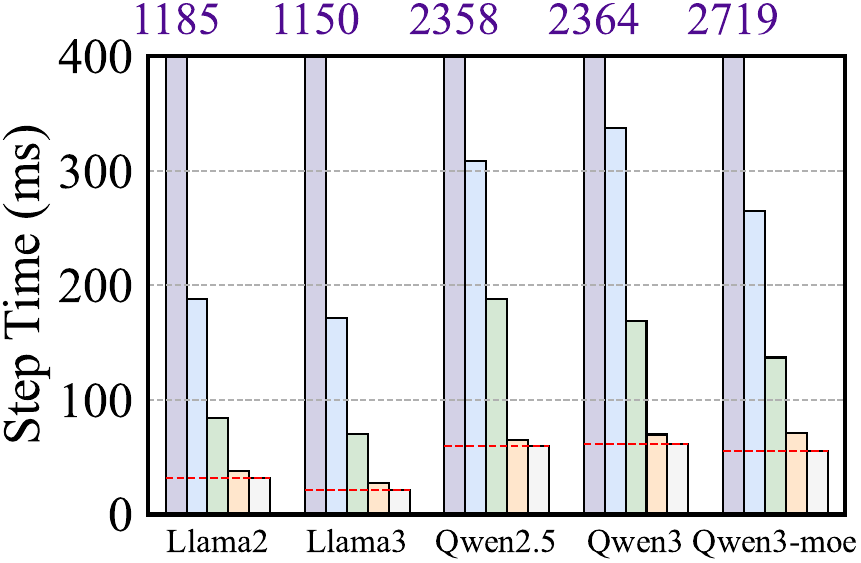}
	\caption{All layers.}
        \label{fig:overlap_benefit_layer}
	\end{subfigure}
 \vspace{-0.1in}
    \caption{Overall transformation cost.}
    \label{fig:overlap_benefit}
    \vspace{-0.3in}
\end{figure}

\vspace{-0.1in}
\subsection{Evaluation Settings}
\vspace{-0.05in}
\parabf{Testbed.}
We have instances of two hardware configurations: H20 and A100. 
Each H20 instance is equipped with eight NVIDIA H20 (96GB) 
GPUs connected via NVLINK, 2 TB of DDR5 memory, and 128 Intel Xeon Platinum 8469C CPUs. 
Each A100 instance is equipped with eight NVIDIA A100 (40GB) GPUs connected via NVLINK, 2 TB of DDR5 memory, and 128 Intel Xeon Platinum 8369B CPUs.

\parabf{Models.}
We employ a mainstream LLMs from Qwen~\cite{Qwen_family} and Llama~\cite{LLaMa_family} families.
The detailed information of selected models is shown in Table~\ref{tab:models}.
The model size ranges from 7B to 32B, which is the dominant and representative size deployed in practical model inference scenarios. 
The selection of GPU instance type for serving each model aligns with the online configuration. 
The fundamental selection strategy is to ensure that a single GPU can accommodate the entire model to achieve optimal performance.

\begin{figure*}[ht!]
    \centering
     \begin{subfigure}{.24\linewidth}
	\centering
	\includegraphics[width=\linewidth]{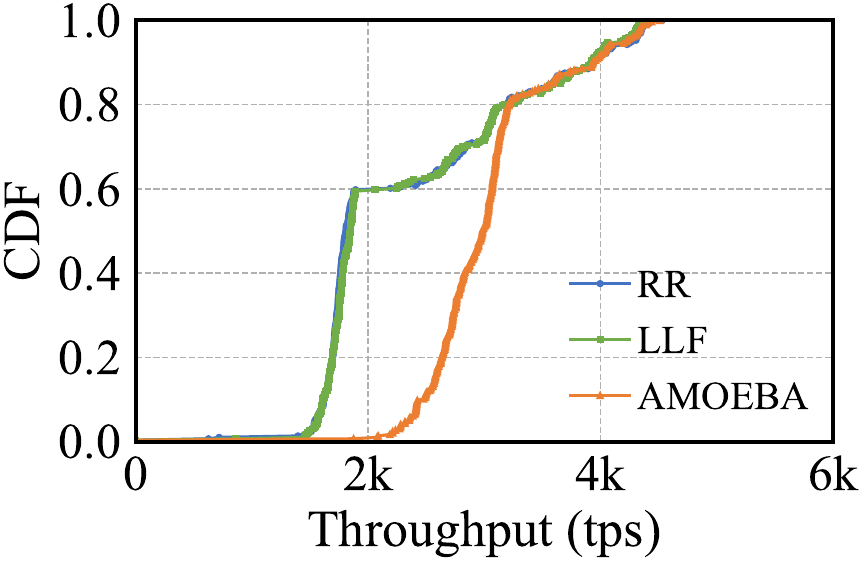}
    \vspace{-0.2in}
	\caption{Llama2.}
        \label{fig:Scheduler_CDF_llama2}
    \end{subfigure}
    \begin{subfigure}{.24\linewidth}
	\centering
	\includegraphics[width=\linewidth]{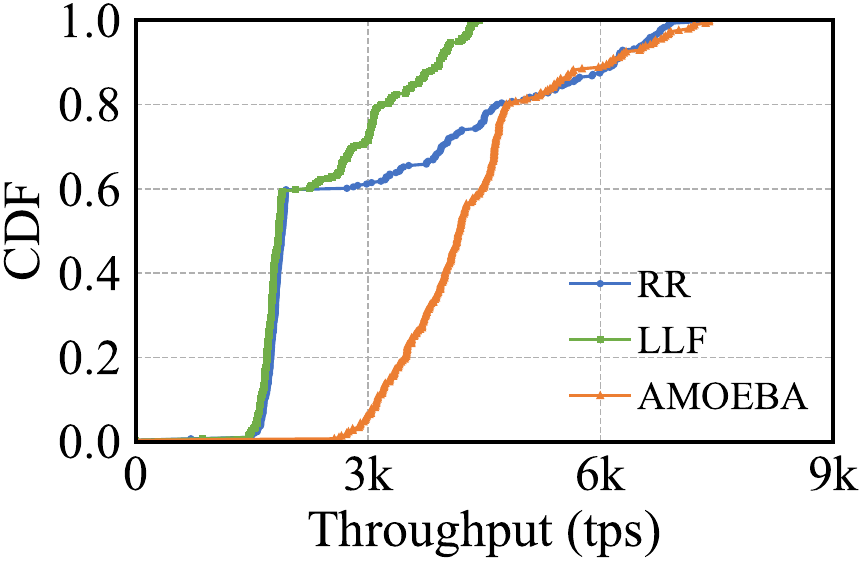}
    \vspace{-0.2in}
	\caption{Llama3.}
        \label{fig:Scheduler_CDF_llama3}
    \end{subfigure}
    \begin{subfigure}{.24\linewidth}
	\centering
	\includegraphics[width=\linewidth]{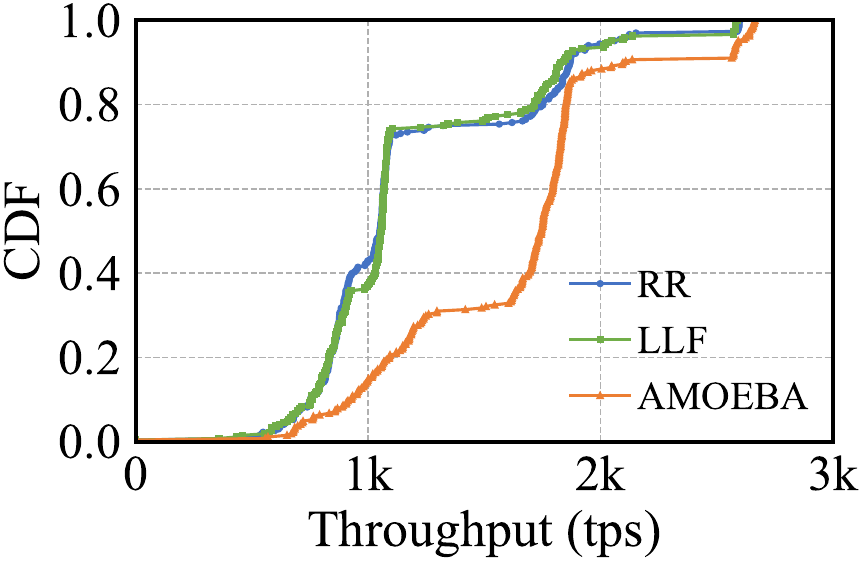}
    \vspace{-0.2in}
    \caption{Qwen2.5.}
        \label{fig:Scheduler_CDF_qwen2.5}
    \end{subfigure}
    \begin{subfigure}{.24\linewidth}
	\centering
	\includegraphics[width=\linewidth]{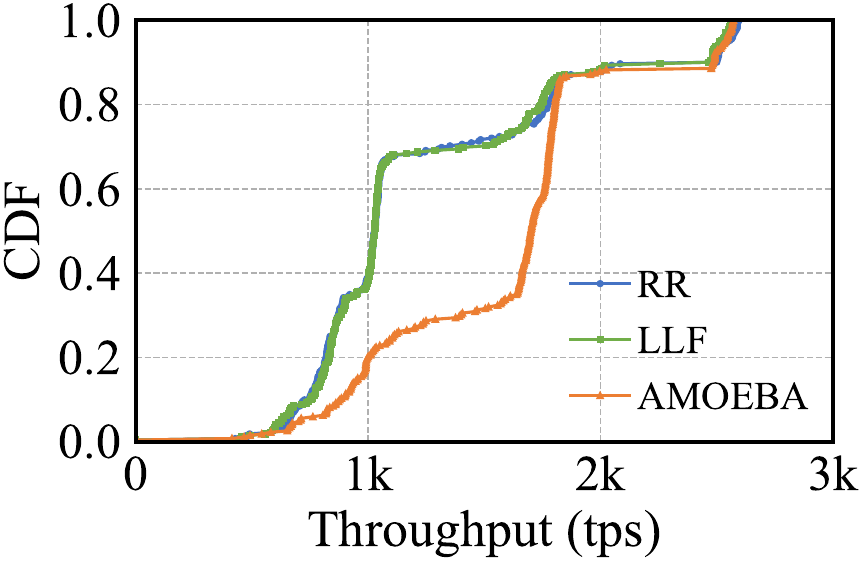}
    \vspace{-0.2in}
    \caption{Qwen3.}
        \label{fig:Scheduler_CDF_qwen3}
    \end{subfigure}
 \vspace{-0.1in}
    \caption{Throughput performance with different scheduling strategies on four distinct models.}
    % \caption{Performance with different scheduling strategies.}
    \label{fig:scheduler}
    \vspace{-0.1in}
\end{figure*}

\begin{figure*}[ht!]
  \centering
  \begin{minipage}[hp]{0.25\linewidth}
    \centering
    \includegraphics[width=\linewidth]{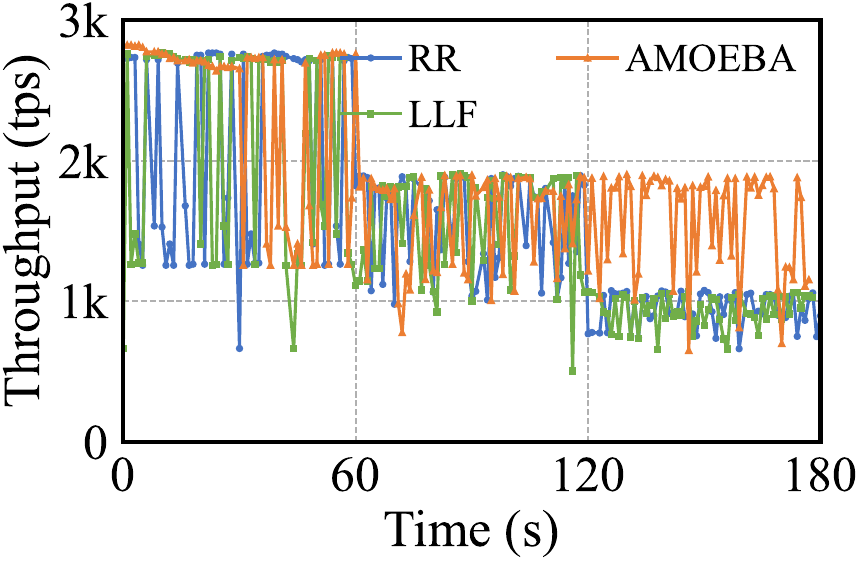}
    \vspace{-0.3in}
    \caption{Throughput Trends}
    \label{fig:Scheduler_sequence}
  \end{minipage}%
  %\hfill
  %\hfill
  \begin{minipage}[t!]{0.75\linewidth}
    \centering
    \begin{subfigure}[t!]{0.32\linewidth}
      \centering
      \includegraphics[width=\linewidth]{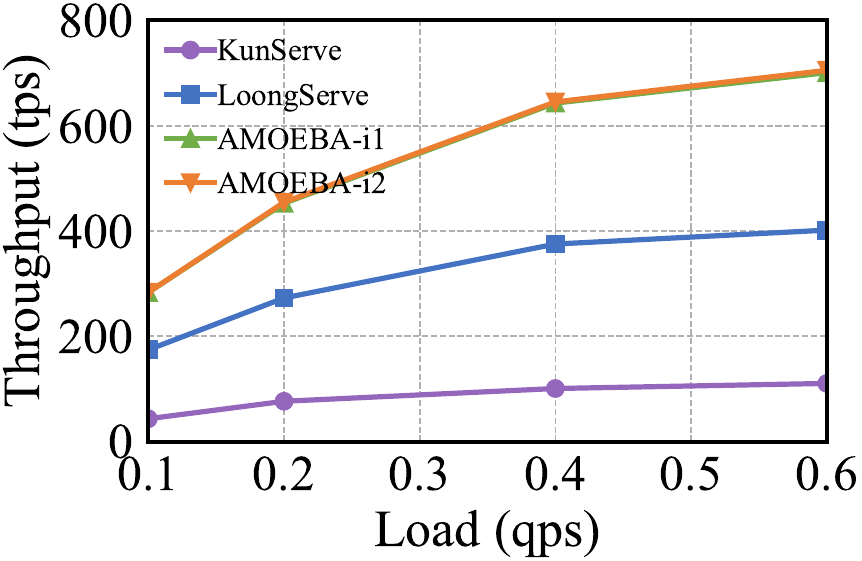}
    \end{subfigure}
    \begin{subfigure}[t!]{0.32\linewidth}
      \centering
      \includegraphics[width=\linewidth]{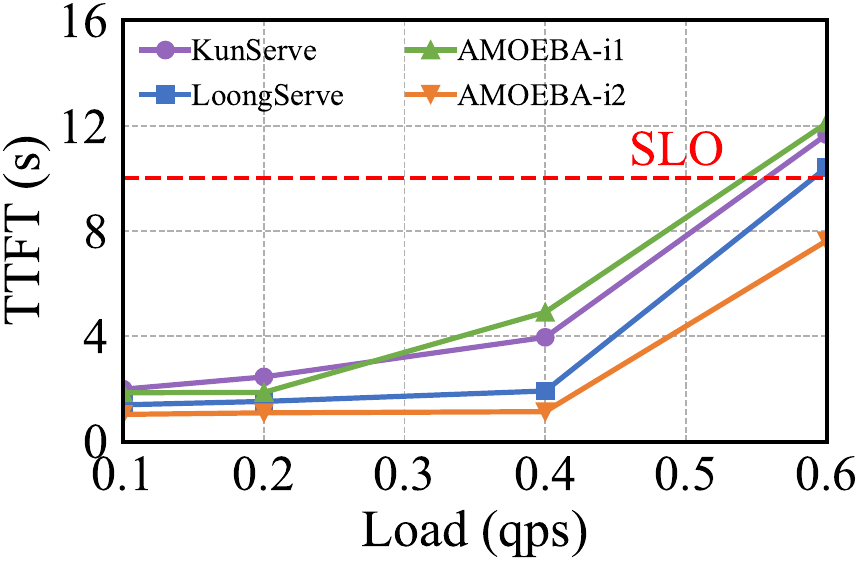}
    \end{subfigure}
    \begin{subfigure}[t!]{0.32\linewidth}
      \centering
      \includegraphics[width=\linewidth]{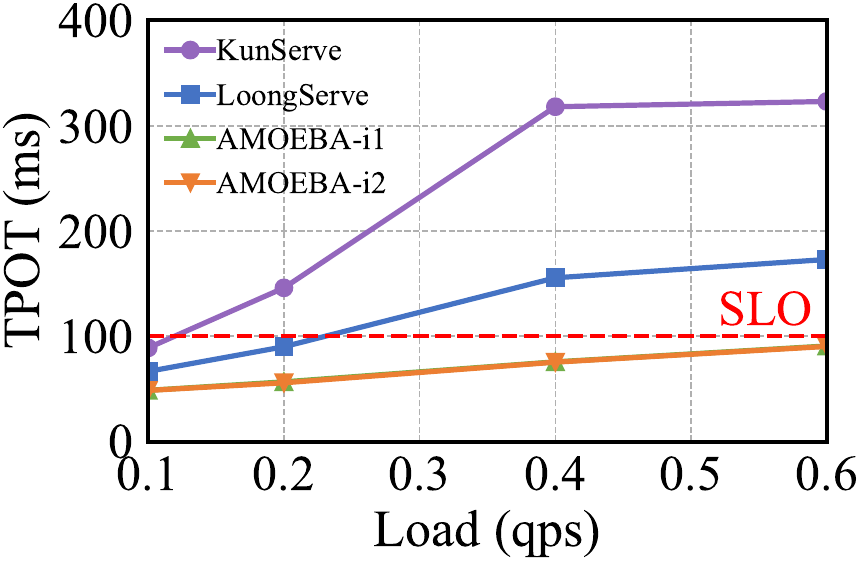}
    \end{subfigure}
    \vspace{-0.12in}
    \caption{End-to-end performance comparison on throughput, TTFT and TPOT.}
    \label{fig:end-to-end}
  \end{minipage}
\vspace{-0.2in}
\end{figure*}

\parabf{Methods in Comparison.} We present \name, along with two improved version, \name-i1 and \name-i2, for evaluation. \name refers to the na\"ive implementation of KV Cache (Migration and De-fragmentation) and weight transformation (Partial Swap). \name-i1 integrates our in-place migration techniques (improvement~1 in \S\ref{sec:subsec:weight_transformation} and \S\ref{sec:scheduler}). \name-i2 further integrates the pipelining techniques (improvement~2 in \S\ref{sec:subsec:weight_transformation} and \S\ref{sec:scheduler}). KunServe \cite{KunServe} and LoongServe \cite{LoongServe} are also included in end-to-end performance comparison as the state-of-the art methods. As KunServe has not open sourced yet, we reproduce it with the online parameter dropping and Network-based KV Cache exchange proposed in the paper.

\vspace{-0.1in}
\subsection{Microbenchmark}
\label{subsec:microbenchmark}
\vspace{-0.05in}
\subsubsection{KV Cache transformation.}
\label{subsubsec:KV}

In this part, we demonstrate the effectiveness of KV Cache transformation.
To clearly compare different methods, we focus on the procedure of a single KV Cache transformation.
%Two solutions are presented here.
% Basic is the basic KV transformation solution presented in \S\ref{subsubsec:kv_migration}.
We use the $4\times(TP1) \to TP4$ as a representative example, which is also the most common transformation for supporting the occurrence of long context requests.
Specifically, for the MoE model (Qwen3-30B-A3B), its transformation is $4\times(TP1EP1) \to TP4EP4$.
Considering that in this scale-up transformation, the KV Cache already reaches high utilization, we set the overall KV Cache utilization to be 90\% in these experiments.
%We varies the used amount of KV cache (from 20\% to 80\% in $TP1$) in different experiments.

\parabf{Transformation time.}
We test the time cost of KV Cache transformation with different solutions in Figure~\ref{fig:KV_cost_evaluation}.
\name introduces 3.15 - 4 ms extra time cost while serving different LLMs, while \name-i1 decreases it by up to 61\%.
%\todo{more details with observation.}
These results show the effectiveness of the page-friendly header-centric KV Cache layout.
With pipelining, \name-i2 further decreases the cost by 86\%.

\parabf{GPU memory saving.}
We further quantify the GPU memory cost and the results are shown in Figure~\ref{fig:KV_memory_evaluation}.
%Mig+Trim inevitably leads to XX\%-XX\% extra memory cost under different LLMs.
%The used memory of \name is 91.6\% less than Mig+Trim.
The memory utilized by \name-i1 is 91.6\% lower than that of \name.
With \name-i1, we consistently maintain additional memory usage below 70 MB, allowing for transformation even under high load scenarios.
%With \name, we can consistently keep the extra used memory to less than 70MB, supporting transformation under extreme high load scenarios.

\vspace{-0.1in}
\subsubsection{Model weights transformation.}
\label{subsubsec:model}

Similarly, we illustrate the performance and overhead of different solutions under a single time of model weights transformation.
% Partial Swap presented in \S\ref{sec:subsec:weight_transformation} represents the basic solution. 
%\name-W represents the optimized solution we proposed in \name.

\parabf{Transformation time.}
The transformation time per layer is shown in Figure~\ref{fig:Model_cost_evaluation}, evaluated on different LLMs.
The time of \name varies from 620 ms to 770 ms, which is mainly caused by the extra migration.
With the weights padding mechanism, \name-i1 completely eliminates unnecessary memory operations, decreasing the cost by 18.9\% - 42.2\%.
With further overlapping, \name-i2 decreases the cost by up to 85\% compared with the \name.

\parabf{Padding overhead.}
% The side-effect introduced by \name-i1 is that we need extra padding in model weights.
We evaluate the extra memory cost and the effect on GEMM computation speed of the extra padding technique in \name-i1
% Therefore, we further evaluate the padding overhead.
We record the memory occupied for model weights in different LLMs, and the results are shown in Figure~\ref{fig:Model_memory_evaluation}.
In various LLMs, the memory overhead ranges from 0\% to 14\%.
Furthermore, we record the corresponding computation time of FFN, before and after padding, for different models.
The extra computation cost is negligible (<0.1\%), validating the effectiveness of the computation-friendly padding in \name-i2.

\vspace{-0.1in}
\subsubsection{Overall transformation cost.}
\label{subsubsec:hybrid}

We further evaluate the performance of \name-i1 under practical execution conditions. The number of layers transformed in each inference step is progressively increased from one to the total number of layers in various models. We make a comparison between Seesaw~\cite{Seesaw}, \name, \name-i1, and \name-i2. 
The results are presented in Figure~\ref{fig:overlap_benefit}.
Raw refers to the original step time without any transformations. 
As the number of transformed layers increases, \name-i1 consistently maintains an overhead of less than 1\%. 
In scenarios where we aim to transform all layers in a single step, \name-i1 reduces the extra cost by 97.2\% compared to Seesaw.

\vspace{-0.1in}
\subsubsection{Transformation-aware scheduler.}
\label{subsubsec:scheduler}

%In this subsection, we evaluate the performance of the transformation-aware scheduler.
In this part, we compare our transformation-aware scheduler with several widely used global scheduling strategies: (1) the Round-Robin (RR) scheduler, which distributes each new request to instances in a round-robin manner; (2) the Least-Load-First (LLF) scheduler, which directs new request to the least-load instance. When an instance is unable to handle a new incoming long request, it collaborates with neighboring instances to implement a scale-up parallelism transformation.
In these experiments, we establish a hybrid workload consisting of both short and long requests. Short requests (1K input length) arrive at a rate of 60 queries per minute, creating the foundational background traffic. 
Long requests (50K input length) arrive at a rate of one query per minute, consistent with the general observations in Figure~\ref{fig:qps}.
%, and each long request has an input length of 50K. 
At initialization, we deploy 8 $TP1$ instances to handle the overall workload. 
%We conduct experiments using different models.
%We take the result of Qwen2.5-32B as a representation, and other results deliver a similar phenomenon.
We sample the throughput periodically during LLM serving and Figure~\ref{fig:Scheduler_CDF_llama2}-\ref{fig:Scheduler_CDF_qwen3} shows the overall CDF of the system throughput performance.
\name improves average throughput by 26.1\%-39.2\% compared to both the RR and LLF. 

Figure~\ref{fig:Scheduler_sequence} captures a representative duration of this experiment.
RR and LLF suffer from excessive transformation of $TP1$ instances when a new long request arrives in the system, while \name delivers higher throughput by better scheduling.
\com{
To highlight the key differences in scheduling, Figure~\ref{fig:Scheduler_sequence} captures a representative duration from this experiment.
At the 120-second time, a $TP4$ instance and $4\times(TP1)$ instances exist, and a new long request arrives. Since the $TP4$ instance is already serving a long request, it experiences a heavier load. As a result, both RR and LLF are more likely to assign the new long request to a $TP1$ instance, prompting another parallelism scale-up and leading to significant performance degradation. In contrast, \name correctly schedules long requests to the existing $TP4$ instance.
}
%The root cause is that with the continuous arrivals of long requests, either RR or LLF would gradually turn all instances into $TP4$, significantly degrading the overall throughput.

\vspace{-0.1in}
\subsection{End-to-End Performance}
\label{subsec:e2e}

We apply the real trace captured in production 
to compare the end-to-end performance of \name-i2 with KunServe and LoongServe under different query-per-second (QPS) settings.
%Besides Seesaw and Basic,
We do not include Seesaw~\cite{Seesaw} since its transformation cost (10s TTFT and 100\% increased TPOT, Figure~\ref{fig:overlap_benefit}) does not satisfy the SLO requirement.
Figure~\ref{fig:end-to-end} shows the result of throughput, TTFT, and TPOT, on Qwen2.5-32B.
Compared with alternatives, \name-i2 increases throughput by 1.75$\times$-6.57$\times$, and TTFT and TPOT are decreased by up to 53\% and 74\%, respectively.
One important gain contributor is that our main design choice (leveraging TP transformation) delivers better performance than PP/SP transformation.
Our pipelining optimization (Improvement~2 in \S\ref{sec:design}) further decreases 26.7\% TTFT.
This optimization plays an important role under the 0.6 QPS load to keep TTFT lower than 10 seconds. 
It enables the system to serve more requests without violating SLO.
%\name improves TTFT of long requests by 16\% compared with the Basic solution, which further validates the effectiveness of the optimizations proposed in this paper.
\vspace{-0.1in}
\section{Conclusion}
\label{sec:conclusion}

To handle the dynamic context lengths and request arrival patterns in LLM serving, we design \name to conduct runtime tensor parallel transformation.
It proposes (1) a page-friendly, header-centric KV Cache layout; (2) dedicated weight padding;  and
(3) a transformation-aware scheduler to comprehensively optimize the performance of serving instances under dynamic workloads.
Evaluations using real-world traces show that \name improves throughput by 1.75$\times$-6.57$\times$ compared to state-of-the-art solutions.
\vspace{-0.1in}
\begin{acks}
This work was supported in part by the Songshan Laboratory Fund (Grant ZZK202402010), the National Natural Science Foundation of China (Project No. 62472101), the Science and Technology Commission of Shanghai Municipality (Project No. 25511107502), and a research grant from Alibaba Group through the Alibaba Innovative Research Program. We thank the anonymous reviewers for their thoughtful feedback.
\end{acks}
\vspace{-0.1in}

\bibliographystyle{ACM-Reference-Format}
\bibliography{amoeba-dac2026}


\end{document}